\documentclass{cernrep} 
\usepackage{texnames}
\usepackage[T1]{fontenc}
\def\smallfrac#1#2{\hbox{${{#1}\over {#2}}$}}
\def\eq#1{Eq.~(\ref{#1})}

\newcommand{\eeq}{\end{equation}}
\newcommand{\bea}{\begin{eqnarray}}
\newcommand{\eea}{\end{eqnarray}}
\newcommand{\la}{\left\langle}
\newcommand{\ra}{\right\rangle}

\def\{{\left\lbrace}
\def\}{\right\rbrace}

\def\frac#1#2{{{#1}\over {#2}}}
\def\gsim{\mathrel{\rlap{\lower4pt\hbox{\hskip1pt$\sim$}}
    \raise1pt\hbox{$>$}}}         
\def\lsim{\mathrel{\rlap{\lower4pt\hbox{\hskip1pt$\sim$}}
    \raise1pt\hbox{$<$}}}         

\newcommand{\new}{\mathrm{new}} 
 
\newcommand{\tot}{\mathrm{tot}}

\newcommand{\draft}[1]{}

\def\Aslash{\not{\hbox{\kern-4pt $A$}}}
\def\Eslash{\not{\hbox{\kern-4pt $E$}}}

\begin{document}
\title{Parton distributions: determining probabilities in a space of functions}
 
\author{{\bf  The NNPDF Collaboration:}
Richard~D.~Ball$^{1}$, Valerio~Bertone$^2$, Francesco~Cerutti$^3$,
 Luigi~Del~Debbio$^1$, Stefano~Forte$^4$, Alberto~Guffanti$^5$, 
Jos\'e~I.~Latorre$^3$, Juan~Rojo$^4$\footnote{Now at PH Department, TH Unit, CERN, CH-1211 Geneva 23, Switzerland} and Maria~Ubiali$^{6}$.}

\institute{~$^1$ School of Physics and Astronomy, University of Edinburgh,\\
JCMB, KB, Mayfield Rd, Edinburgh EH9 3JZ, Scotland\\
~$^2$  Physikalisches Institut, Albert-Ludwigs-Universit\"at Freiburg,\\ 
Hermann-Herder-Stra\ss e 3, D-79104 Freiburg i. B., Germany  \\
~$^3$ Departament d'Estructura i Constituents de la Mat\`eria, 
Universitat de Barcelona,\\ Diagonal 647, E-08028 Barcelona, Spain\\
~$^4$ Dipartimento di Fisica, Universit\`a di Milano and
INFN, Sezione di Milano,\\ Via Celoria 16, I-20133 Milano, Italy\\
~$^5$ The Niels Bohr International Academy and Discovery Center,\\
The Niels Bohr Institute, Blegdamsvej 17, DK-2100 Copenhagen, Denmark\\
~$^6$ Institut f\"ur Theoretische Teilchenphysik und Kosmologie, RWTH Aachen University,\\ 
D-52056 Aachen, Germany}

\maketitle 

\begin{abstract}
We discuss the statistical properties of parton distributions 
within the framework of the NNPDF methodology. 
We present various tests of statistical consistency, in particular that the
distribution of results does
not depend on the underlying parametrization and that it behaves
according to
Bayes' theorem upon the addition of new data. We then study the
dependence of results on consistent or inconsistent datasets and
present tools to assess the consistency of new data. Finally
we estimate the relative size of the PDF uncertainty due to  data uncertainties, and that due to the  
need to infer a functional form from a finite set of data.
\end{abstract}
 
\section{The NNPDF approach to parton distributions}

The determination of parton distributions (PDFs) and their
uncertainties~\cite{Forte:2010dt} 
poses a difficult problem because one is trying to
determine the probability distribution for a set of functions. Given
that this is necessarily done from a finite set of data it requires
some assumptions: some of these, such as a certain degree of
smoothness, may be  physically  motivated, but it is important to
check that they do not bias the result and in particular that they do
not destroy its statistical interpretation. The most common way of
implementing these assumptions is to assume a functional form for the
PDFs, each parametrized by a small number of parameters  (typically
between two and five)  which are determined by fitting a suitable set of
data. The NNPDF collaboration has developed an alternative 
approach~\cite{Forte:2002fg,DelDebbio:2004qj,DelDebbio:2007ee,Ball:2008by,Rojo:2008ke,Ball:2009mk,Ball:2010de,Ball:2011mu}
which tries to avoid the bias associated to  this procedure.

The NNPDF approach is based on four main ingredients:
\begin{itemize}
\item {\it Monte Carlo by importance sampling.} NNPDF produces a  Monte
  Carlo sampling of the probability density in the (function) space of
  PDFs. To adequately sample this space by simple binning would be simply 
 impossible: for example assuming seven PDFs
  (the three light quarks and antiquarks and the gluon) sampled at ten
  points, binning the probability distribution in each direction
  with five bins one would end up with $5^{70}\sim 10^{49}$ bins. The
  problem is solved by importance sampling: most bins are empty and
  only those with data are relevant. Hence, one starts by constructing
  a set of data replicas, which reproduces the statistical features of
  the original data. It then turns out that a sample of 1000
  pseudo-data replicas is large enough 
to reproduce central values, uncertainty and correlations
  of the starting data to a few percent accuracy
\item {\it Neural networks as universal unbiased 
interpolants.} Each of the underlying functions is parametrized with a
feed-forward multilayer neural network. The architecture chosen
corresponds to 37 free parameters for each of the  seven PDFs. It can
then be checked that results do not depend on the parametrization by
verifying that they are unchanged if the size of the neural network
is reduced. 
\item {\it Genetic Algorithms for neural network training.} The best fit is
  determined by using  a genetic algorithm, and starting from a
  random initialization of parameters. This ensures that the
  presumably wide
space of equivalent minima can be adequately explored.
\item {\it Determination of the best fit by 
cross-validation.} Because the parametrization is very large, the best
fit is not the minimum of the $\chi^2$, which would correspond to
fitting noise. The best fit is then found by dividing randomly data in two
sets (training and validation) for each experiment, minimizing the
$\chi^2$ of the training set while monitoring the $\chi^2$ of both
sets. The best-fit is obtained when the $\chi^2$ 
of the validation set starts increasing despite the fact that the
$\chi^2$ of the training set still decreases.
\end{itemize}

\section{Statistical consistency}
\label{sec:cons}

Our starting point is the NNPDF2.1 NLO~\cite{Ball:2011mu} PDF set: we
would like to test that it behaves in a statistically consistent way.
For a start, in Table~\ref{tab:estfit1} we show the statistical
estimators for this PDF fit: $\chi^{2}_{\tot}$ is the result of the
comparison to data of the best-fit PDFs (defined as the average over the
$N_{\rm rep}=1000$ replicas of the Monte Carlo sample); $\la
\chi^{2(k)} \ra$ is the average of the values obtained by comparing
each PDF replica to the data, and   $\la E \ra$ is the value of the same
figure of merit, but obtained obtained comparing  each PDF replica to
the corresponding data replica. For the latter, the training and
validation values are also shown. All figures of merit are computed
using the full covariance matrix, with normalization uncertainties
included using the so-called $t_0$ method of Ref.~\cite{Ball:2009qv};
they are all normalized to the number of data points $N_{\rm dat}$.
The fact that $\la
\chi^{2(k)} \ra\sim 1$ while  $\la E \ra\sim 2$, and also that
$\chi^2_{\rm tot}<\la
\chi^{2(k)} \ra$ are both  consistent with the
fact that the fit is ``learning'' an underlying law: the fitted PDFs
are closer to the data than the data replicas (despite being fitted to
the latter), and the best fit (obtained averaging replicas) is yet
closer to the data than any of the individual replicas.
\begin{table}[h!]
\centering
\small

\begin{tabular}{|c||c||c||c|}
\hline
  & Reference & Central Values & Average Fixed Partitions \\
\hline
\hline 
$\chi^{2}_{\tot}$ &      1.16  &  1.14  &  1.15 \\
$\la E \ra \pm \sigma_{E} $   &     $2.24\pm 0.09$  & $1.25\pm 0.11$ &
$1.24\pm 0.07$   \\
$\la E_{\rm tr} \ra \pm \sigma_{E_{\rm tr}}$&   $2.22 \pm 0.11$    &
$1.25 \pm 0.12$ &  $1.23\pm 0.07$  \\
$\la E_{\rm val} \ra \pm \sigma_{E_{\rm val}}$&   $2.28 \pm 0.12$     & 
$1.27\pm 0.11$&  $1.26\pm 0.08$   \\
\hline
$\la \chi^{2(k)} \ra \pm \sigma_{\chi^{2}} $  &  $1.25\pm 0.09$  &
$1.25\pm 0.11$ &  $1.24\pm 0.07$  \\
\hline
\end{tabular}
\caption{\small \label{tab:estfit1} Table of statistical estimators
  for NNPDF2.1 with $N_{\rm rep}=
1000$ replicas (first columns).  The subsequent columns show the
corresponding results, to be discussed in Sect.~\ref{sec:funcdat}, for
fits to central data and with
fixed partitions,  with  $N_{\rm rep}=
100$ replicas each. All entries in the last column are obtained
repeating the procedure for five random choices of fixed partition and
averaging the final results. All values are normalized to the
number of data points.}

\end{table}

More detailed tests can be performed by looking at the distance between
estimators extracted from  PDF sets, defined as follows.
Given a set of $N^{(i)}_{\mathrm{rep}}$ 
PDF replicas, the estimator for any quantity $q$ computed from the PDFs
(including the PDFs themselves) is
  the mean
$    \langle
    q\rangle_{(i)}=\frac{1}{N^{(i)}_{\mathrm{rep}}}\sum_{k=1}^{N^{(i)}_{\mathrm{rep}}}q_k$. The
    distance between two determinations of $q$  from sets
  $q^{(1)}_i$, $q^{(2)}_i$  
  is then
  \begin{equation}
    \label{eq:d2}
    d^2\left(\langle q^{(1)}\rangle ,\langle q^{(2)}\rangle\right)=
    \frac{\left(\langle q^{(1)}\rangle_{(1)} - \langle q^{(2)}\rangle_{(2)}\right)^2}
    {\sigma^2_{(1)}[\langle q^{(1)}\rangle] +
      \sigma^2_{(2)}[\langle q^{(2)}\rangle]},
  \end{equation}
  with the variance of the mean given by
\begin{equation}\label{eq:varmean}
     \sigma^2_{(i)}[\langle q^{(i)}\rangle]=
    \frac{1}{N^{(i)}_{\mathrm{rep}}} \sigma^2_{(i)}[q^{(i)}]
\end{equation}
  in terms of the variances  $\sigma^2_{(i)}[q^{(i)} ]$ of the variables
  $q^{(i)}$ (which a priori could come from two distinct probability
  distributions). The distance between uncertainties can be defined in
  a similar way.
By construction, the probability distribution for the distance coincides
with the $\chi^2$ distribution with one degree of freedom, and thus it
has mean $\langle d\rangle= 1$, and $d\lsim 2.3$ at 90\% confidence
level. 

\begin{figure}
\centering\includegraphics[width=.99\linewidth]{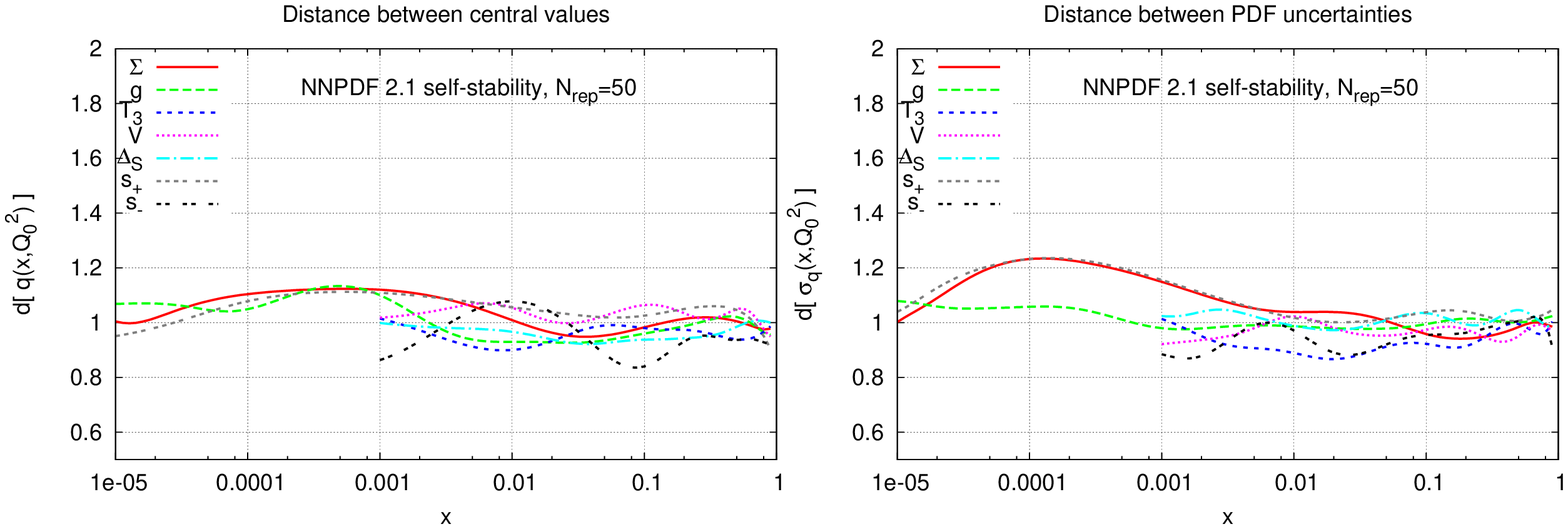}
\includegraphics[width=.99\linewidth]{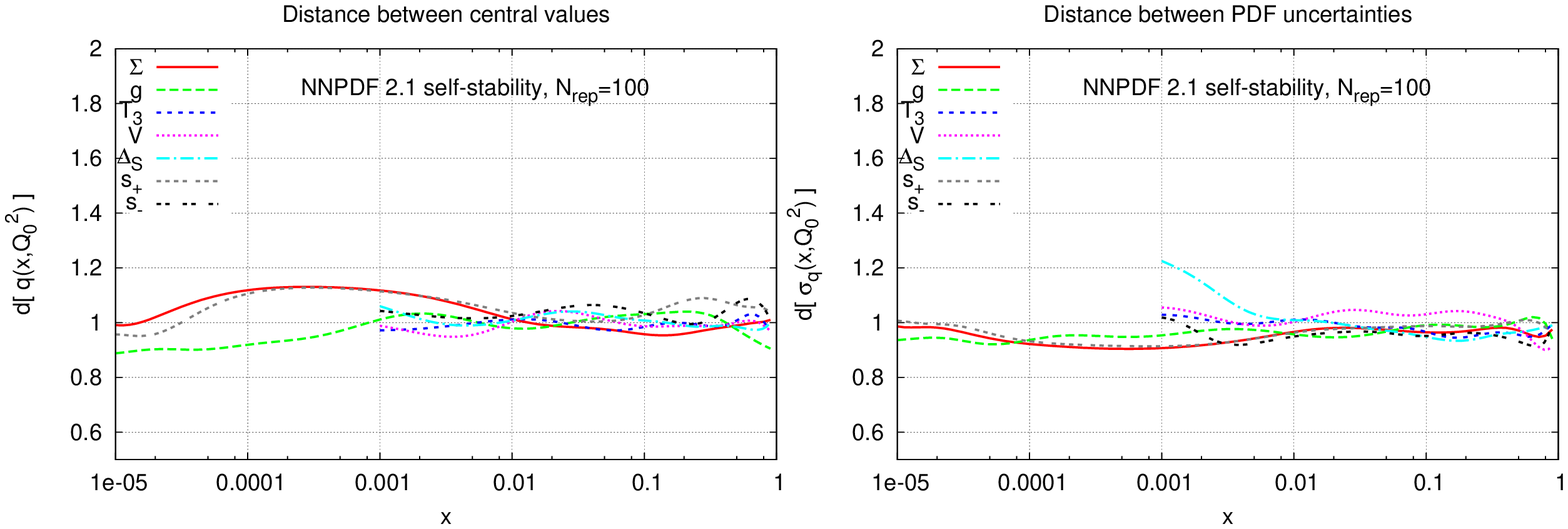}
\caption{Distances between central values and uncertainties of
PDFs computed from two distinct  sets of 
$N_{\rm rep}=50$  (top) or
$N_{\rm rep}=100$ replicas (bottom).}
\label{fig:nnpdf-distances-selfstab}
\end{figure}

An immediate use of the distance is to check that PDF sets computed
from different sets of replicas are statistically equivalent
(i.e. that $\langle
    q^{(k)}\rangle_{(i)}$ has the expected distribution). This is
    shown in  Fig.~\ref{fig:nnpdf-distances-selfstab} (top row): 
indeed distances
    fluctuate about $d\sim1$. Furthermore, one can check
    (Fig.~\ref{fig:nnpdf-distances-selfstab}, bottom row) that the
 distance does not change as the number of replicas is varied: because
of the explicit factor of $\frac{1}{N^{(i)}_{\mathrm{rep}}}$ in
Eq.~(\ref{eq:varmean}), this verifies that indeed the uncertainty of
the mean decreases as $1/\sqrt{N^{(i)}_{\mathrm{rep}}}$ as
$N^{(i)}_{\mathrm{rep}}$ is increased. Note that this means that the distance
between two PDFs that barely overlap within error bands at 
68\% C.L. with
$N_{\rm rep}=100$ replicas is $\langle d\rangle\sim 7$  (because the 
distance is computed averaging results from  subsets  of
$N_{\rm rep}/2=50$ replicas~\cite{Ball:2009mk}).

\begin{figure}
\centering\includegraphics[width=.99\linewidth]{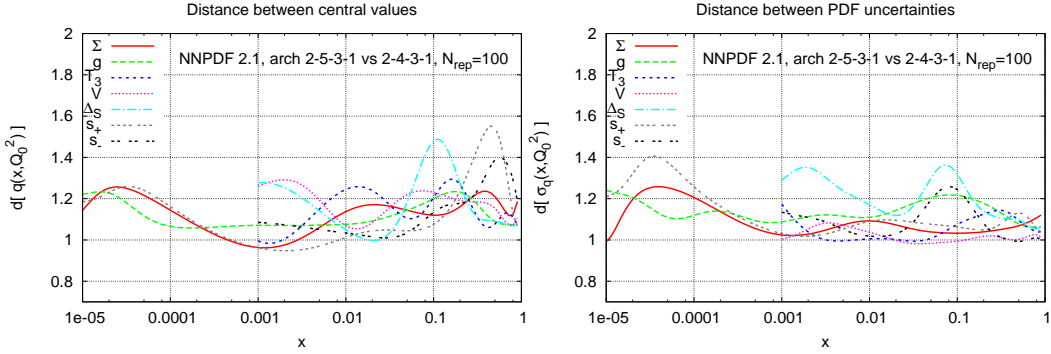}
\caption{Distances between PDFs with the default neural network
  architecture
(2--5--3--1) and  a reduced  architecture (2--4--3--1).}
\label{fig:nnpdf-distances-arch}
\end{figure}

Next, we  check the independence of results of the
parametrization. This is done by constructing a new set of PDF
replicas with a different choice of architecture for neural networks,
and checking that results are statistically equivalent. In 
Fig.~\ref{fig:nnpdf-distances-arch} we show the  distances 
between PDFs based on the default architecture 2--5--3--1.
and PDFs based on the smaller 2--4--3--1 architecture. 
This corresponds to removing 6 free parameters from the
parametrization of each PDF, i.e. removing
of 42 free parameters overall. The similarity of
Figs.~\ref{fig:nnpdf-distances-selfstab}
and~\ref{fig:nnpdf-distances-arch} proves the stability of
results. Note that, in order to make sure that the parametrization is
indeed redundant, the larger architecture is used as a default.

Finally, we turn to our most detailed test of the statistical
consistency of PDFs determined with the NNPDF methodology.
Namely, we exploit the fact that given the probability distribution
${\mathcal P}_{\rm old}(f)$ 
for PDFs determined from a certain starting dataset, the effect of the 
inclusion  of the information from new data can be determined using
Bayes' theorem. It is then possible to compare the probability
distribution ${\mathcal P}_{\rm new}(f)$  obtained in
this way, with a determination of ${\mathcal P}_{\rm new}(f)$ 
 found by simply performing a fit to an extended
dataset including both the starting dataset and the new
data. Statistical equivalence of the
two determinations of  ${\mathcal P}_{\rm new}(f)$ shows that the
NNPDF methodology treats the information contained in the data in a
consistent way. In fact, repeating this test for all of the data used for the
fit, to the extent that for a large enough dataset
results are independent of the prior assumption,  
would amount to a proof that the set of data and the set of PDFs
determined from it contain the same information (``closure test''): 
indeed, such a
Bayesian procedure was suggested in Ref.~\cite{Giele:2001mr} as a
way of arriving at a fully unbiased and self-consistent PDF
determination.

We have performed such a  test for an
individual subset of data included in the NNPDF2.1 NLO PDF
determination. The formalism to do so was developed in
Ref.~\cite{reweighting,Ball:2011gg}, correcting a previous proposal of
Ref.~\cite{Giele:1998gw}. 
The way it works is the following: assume we want to include $n$ new
data 
$
  y = \left\lbrace y_1,y_2,\cdots,y_n \right\rbrace 
$
which had not been originally included 
in the determination of the initial probability density
distribution. We view this data as a point $y$
in an $n$-dimensional space, with  uncertainties given as
a  $n\times n$ experimental
covariance matrix.
We update the probability density
$\mathcal{P}_{\rm old}(f)$ using the conditional probability 
of the new data, which is proportional to the probability density of 
the $\chi^2$ to the new data conditional on $f$:
\begin{equation}
  \mathcal{P}(\chi^2|f) \propto (\chi^{2}(y,f))^{{1\over 2}(n-1)}
  e^{-\frac{1}{2}\chi^{2}(y,f)},
  \label{eq:csqdist}
\end{equation}
where $y_i[f]$ is the value predicted for the data
$y_i$ using the PDF $f$.
By Bayes' theorem then
\begin{equation}
\mathcal{P}_{\rm new}(f)
= \mathcal{N}_{\chi}\mathcal{P}(\chi|f)\;\mathcal{P}_{\rm old}(f),
\label{eq:multiplication}
\end{equation}
(with $\mathcal{N}_{\chi}$ an $f$--independent
normalization factor). 

Using Eq.~(\ref{eq:csqdist}) in Eq.~(\ref{eq:multiplication})
immediately implies that the inclusion of the new data can be viewed
as a reweighting of the prior probability distribution
$\mathcal{P}_{\rm old}(f)$.
Namely, if the expectation value of some observable  $\mathcal{O}$ with 
the distribution $\mathcal{P}_{\rm old}(f)$
is \begin{equation}
\la\mathcal{O}\ra=\smallfrac{1}{N}\,\sum_{k=1}^{N}
\mathcal{O}[f_k]\, ,
\label{eq:avg} 
\end{equation}
then, by Eq.~(\ref{eq:multiplication}), 
its expectation value according to $\mathcal{P}_{\rm old}(f)$ is
\begin{eqnarray}\label{eq:avgnew}
\la\mathcal{O}\ra_{\new}&=&\smallfrac{1}{N}\,\sum_{k=1}^{N}
\mathcal{N}_\chi\mathcal{P}(\chi|f_k)
\mathcal{O}[f_k] 
=\smallfrac{1}{N}\sum_{k=1}^{N} w_k \, \mathcal{O}[f_k]\,
,
\end{eqnarray}
with
\begin{equation}
w_k = 
\frac{(\chi^{2}_k)^{{1\over 2}(n-1)} 
e^{-\frac{1}{2}\chi^{2}_k}}
{\smallfrac{1}{N}\sum_{k=1}^{N}(\chi^{2}_k)^{n/2-1}
e^{-\frac{1}{2}\chi^{2}_k}}\, .
\label{eq:weights}
\end{equation}
The weights $w_k$, when divided by $N=N_{\rm rep}$, are just the 
probabilities of the replicas $f_k$, given the $\chi^2$ to the new data. 

  \begin{figure}[t!]
    \begin{center}
      \includegraphics[width=0.45\textwidth]{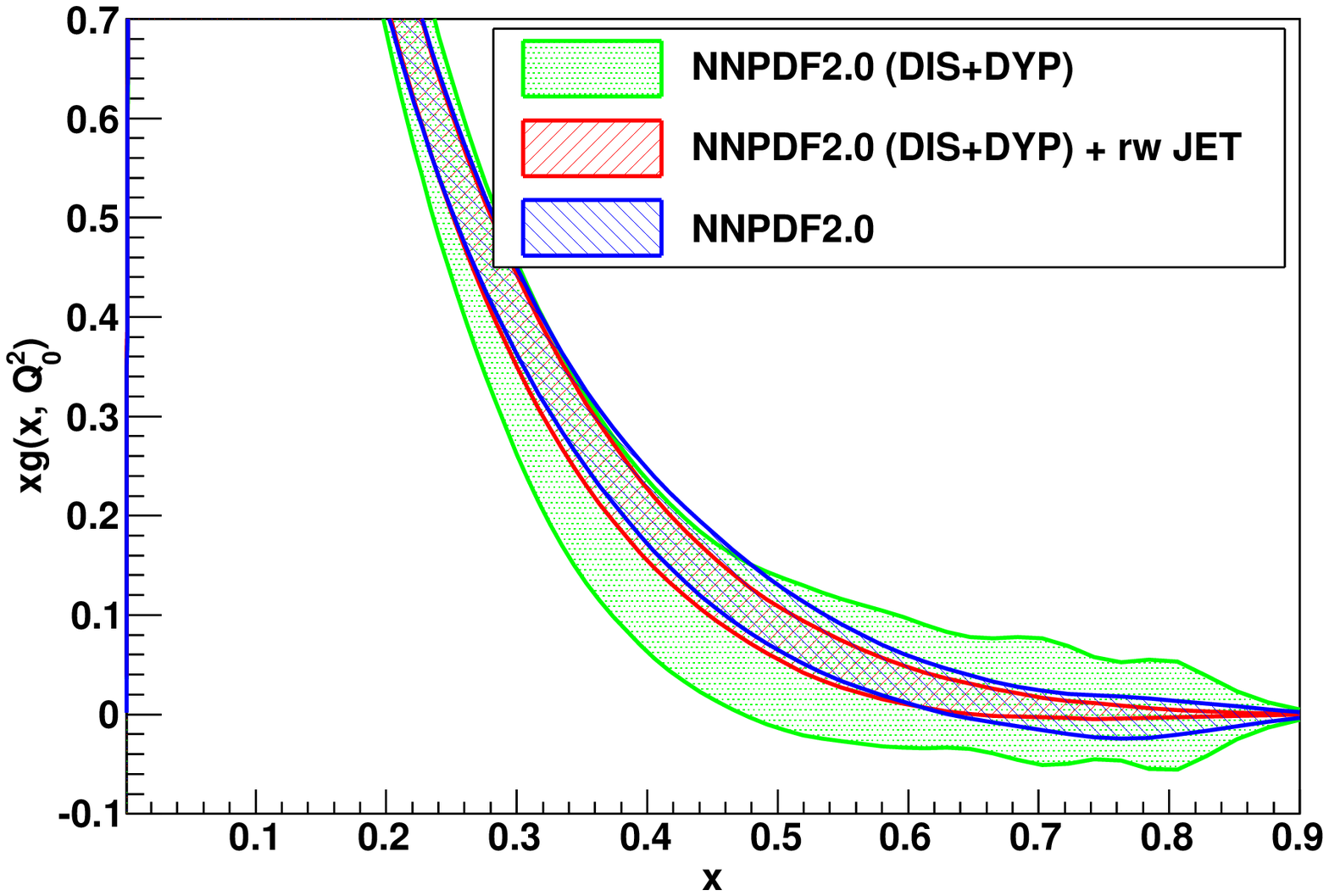}
      \includegraphics[width=0.45\textwidth]{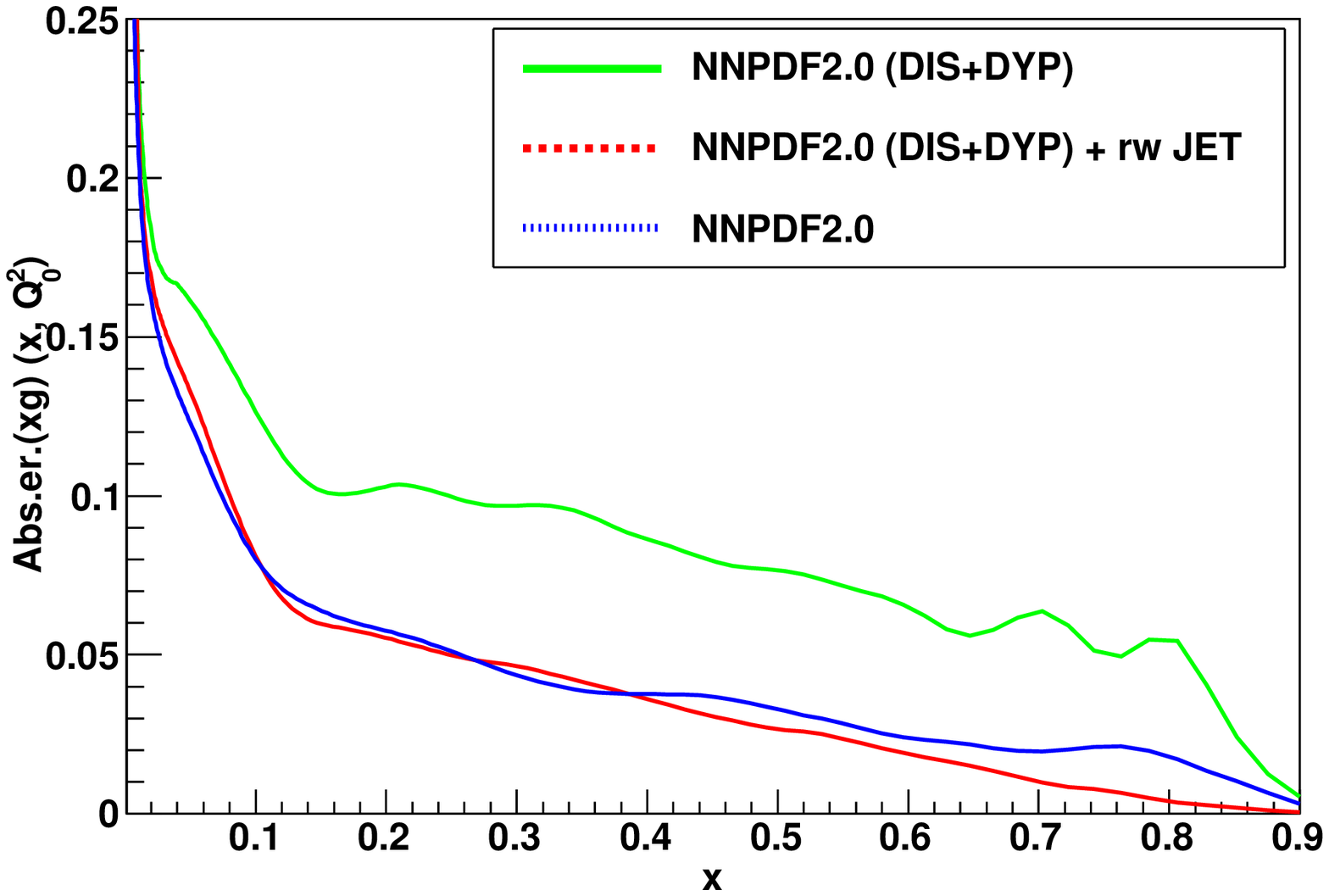}\\
      \includegraphics[width=0.45\textwidth,clip]{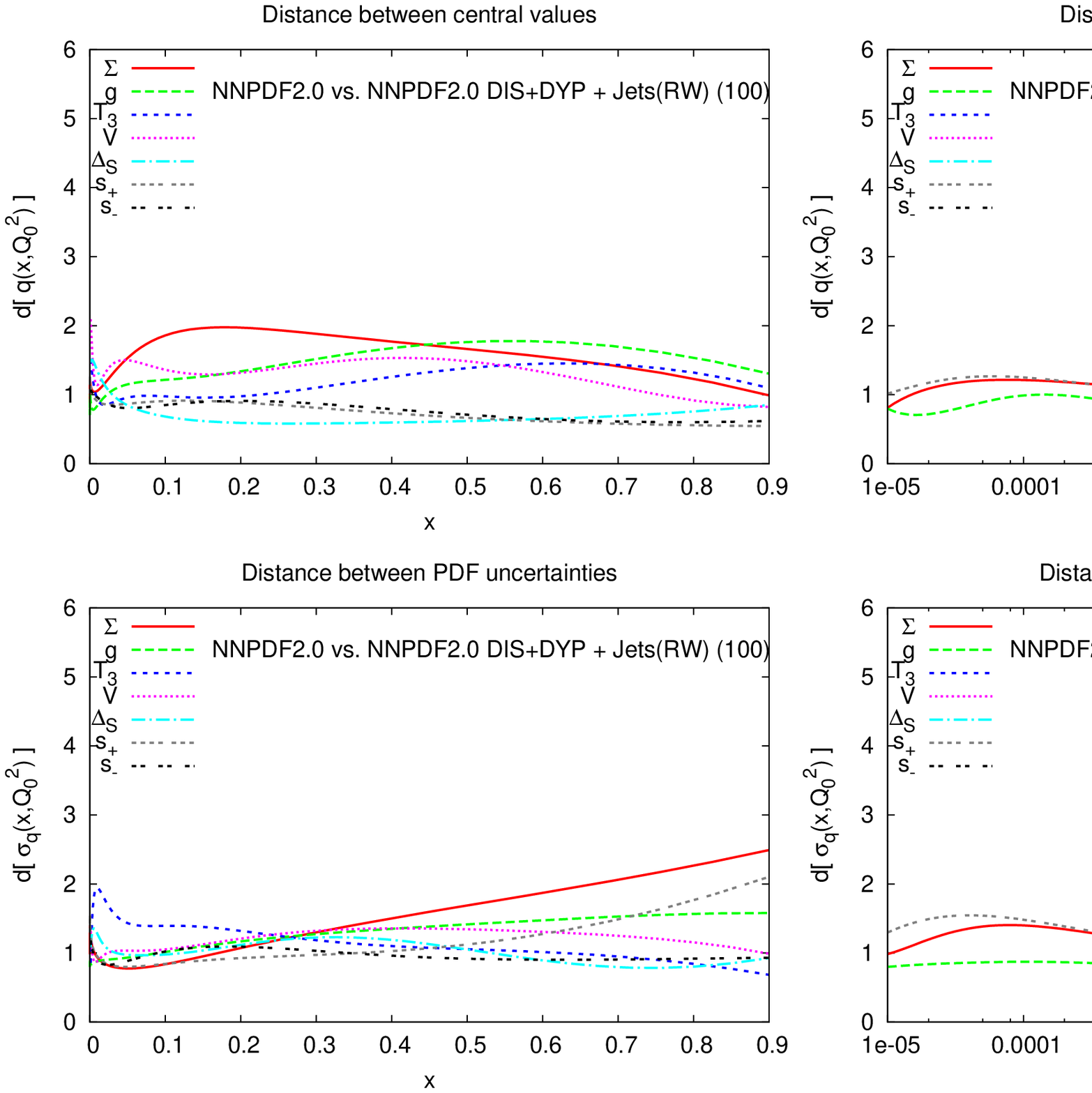}
      \includegraphics[width=0.45\textwidth,clip]{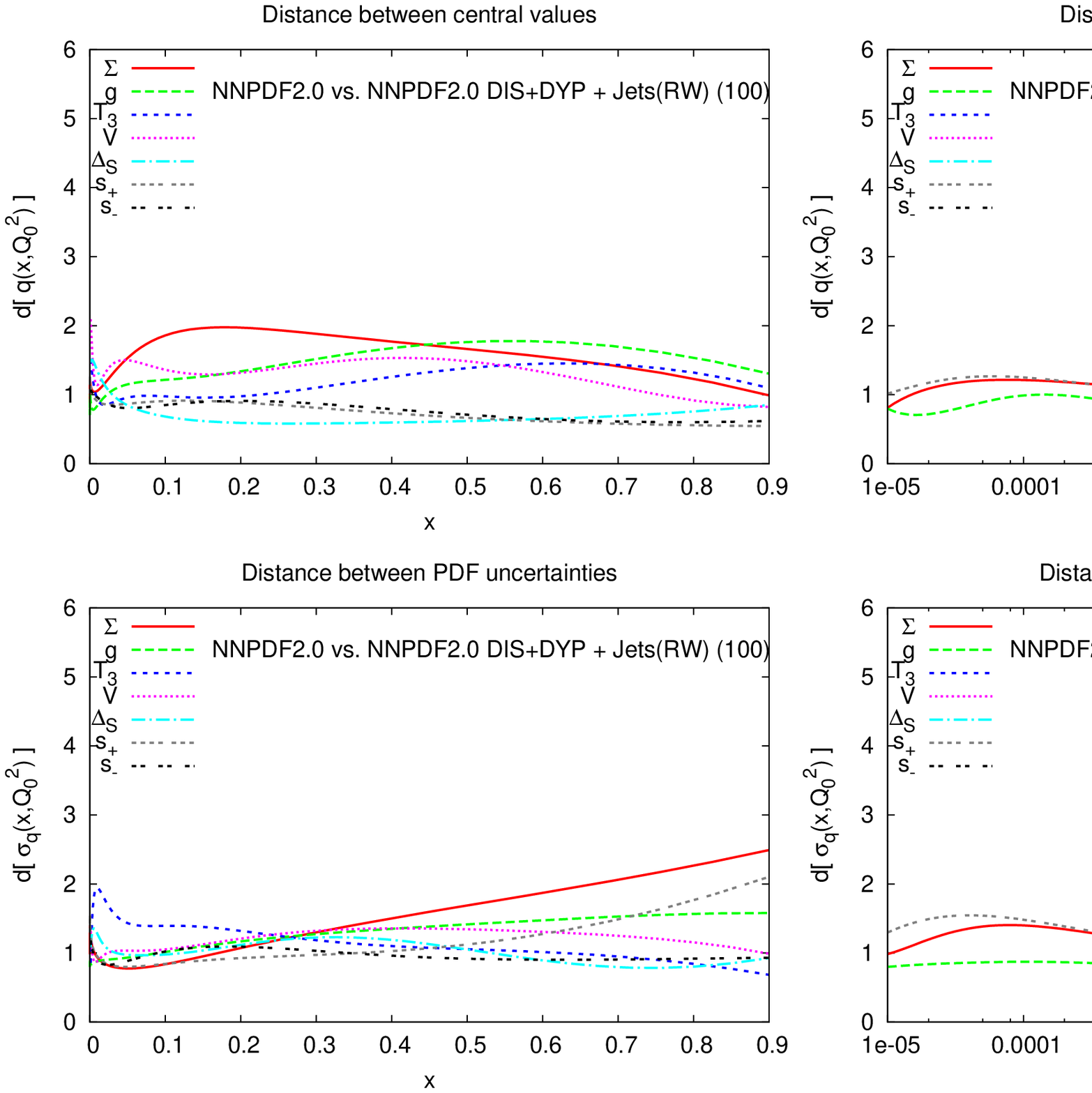}
    \end{center}
    \vskip-0.5cm
    \caption{\small Top: the gluon distribution (left) and its uncertainty (right) of 
      the NNPDF2.0(DIS+DY) fit before and after reweighting with the inclusive jet data 
      compared to the refitted gluon from NNPDF2.0. Bottom: distances
      between the refitted and reweighted results for central values
      (left) and uncertainties (right).}
    \label{fig:pdf-jets}
\end{figure}

The comparison between the ``reweighted'' result
Eq.~(\ref{eq:avgnew}-\ref{eq:weights}) and the refitted one is shown
in Fig.~\ref{fig:pdf-jets}:  it is apparent that the two procedures lead
to the same result, except possibly at very large $x\gsim 0.7$ where
the determination becomes unreliable because of the lack of
experimental information. This is a very strong check that PDF
uncertainties admit a {\it bona fide} statistical interpretation, and
thus should not be viewed of theoretical uncertainties with unknown
distribution. Note that because NNPDF results are delivered as a Monte
Carlo sample, any feature of the distribution of results, such as
confidence intervals or higher moments, can be determined explicitly.

\section{Dataset dependence}

One important feature of the NNPDF approach  
is that the same methodology can be used to determine PDFs from
datasets of rather different size and nature: this, in particular,
follows from the extreme redundancy of the parametrization, and the
ensuing parametrization independence, explicitly checked in the
previous section.  In fact, NNPDF results are even stable
upon the addition of new independently parametrized  PDF, as seen in
Ref.~\cite{Rojo:2008ke,Ball:2008by} where light quark and gluon PDFs
were found to be stable upon addition of an independent parametrization
of strangeness. 
This is to be contrasted to the approach used by
other groups, where a larger dataset requires the introduction of 
more parametrs. As a consequence, in the NNPDF approach, unlike in
other approaches, the addition of new compatible data results in error
reduction, as has been checked explicitly in benchmark 
studies~\cite{Ball:2008by,Dittmar:2009ii}. 

\begin{figure}[t!]
  \begin{center}
    \includegraphics[width=0.45\textwidth]{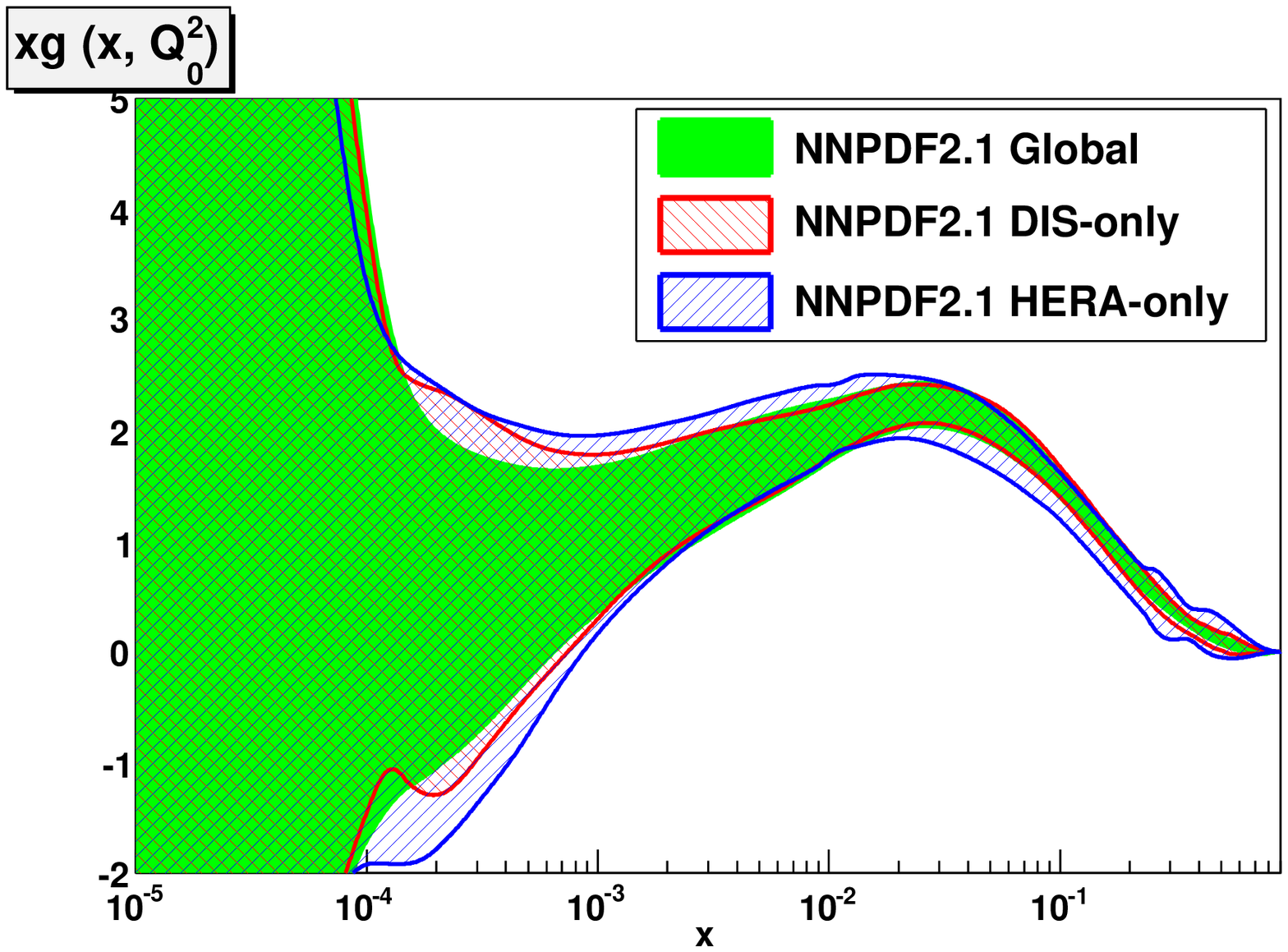}
\includegraphics[width=0.45\textwidth]{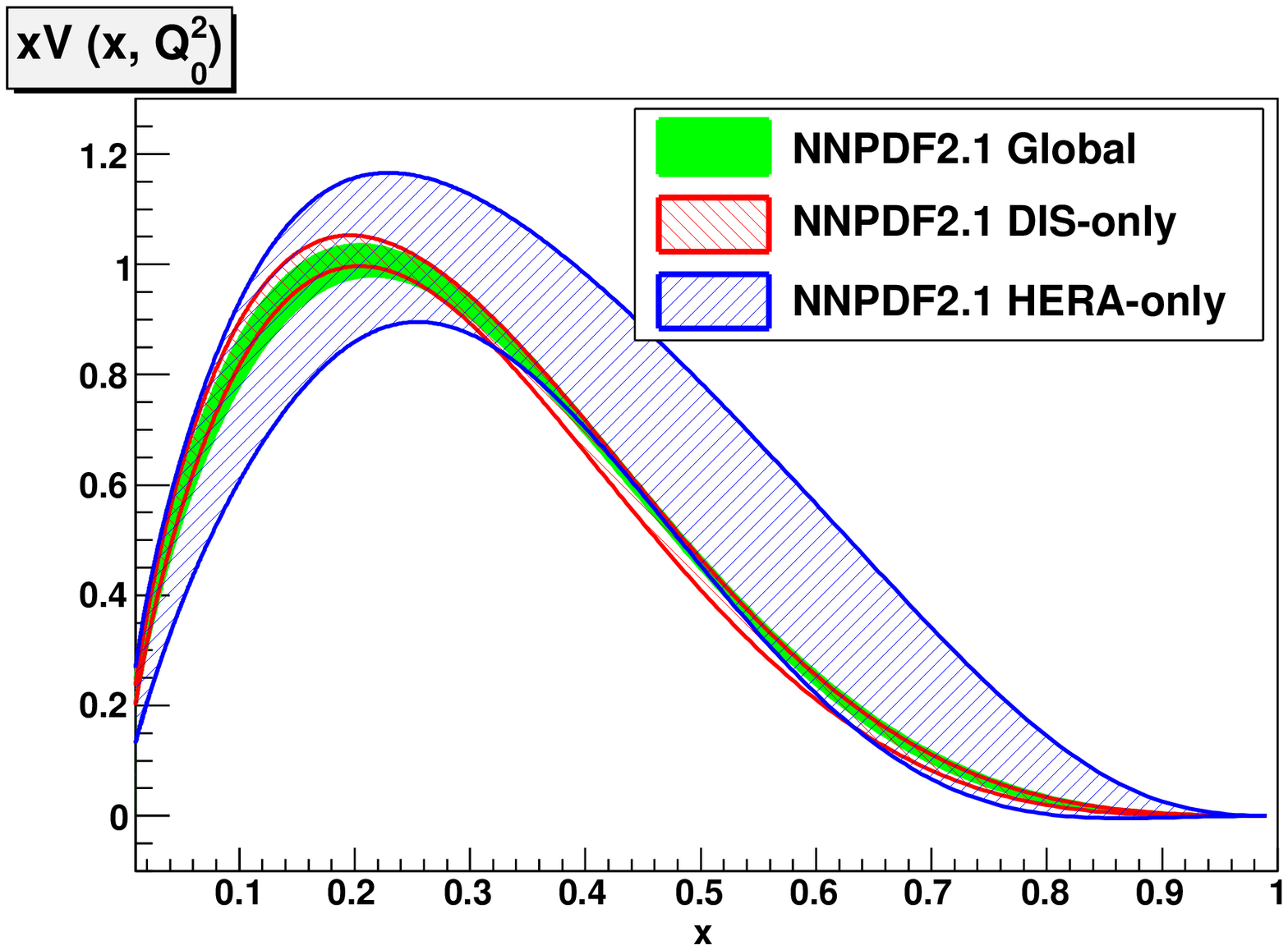}
  \end{center}
  \caption{\small Comparison of PDFs obtained to fits
to different datasets: a global fit, a DIS--only fit and
a HERA--only fit. The gluon (left) and total
valence (right) PDFs  are shown.
\label{fig:PDFs-reduced} }
\end{figure}

By comparing the results of fits to different datasets it is then
possible to study the effect of individual data on PDFs and verify
their consistency. For example, in  
Fig.~\ref{fig:PDFs-reduced}, we compare the default NNPDF2.1 PDF set
to PDFs obtained using only the DIS data or only the HERA DIS data
from the global dataset. On the one hand, it is apparent from this
comparison that these fits are mutually consistent; on the other hand
it is clear that the HERA data determine well the small $x$ gluon, the
DIS data also determine well the total valence (mostly due to neutrino
data), while the global dataset further improves the large $x$
gluon. Detailed studies of this kind are performed in
Refs.~\cite{Ball:2010de,Ball:2011mu} (see Ref.~\cite{Forte:2010dt} for
a general discussion of  the expected impact of different
data on PDFs).

\begin{figure}[h!]
\begin{center}
\includegraphics[width=0.48\textwidth]{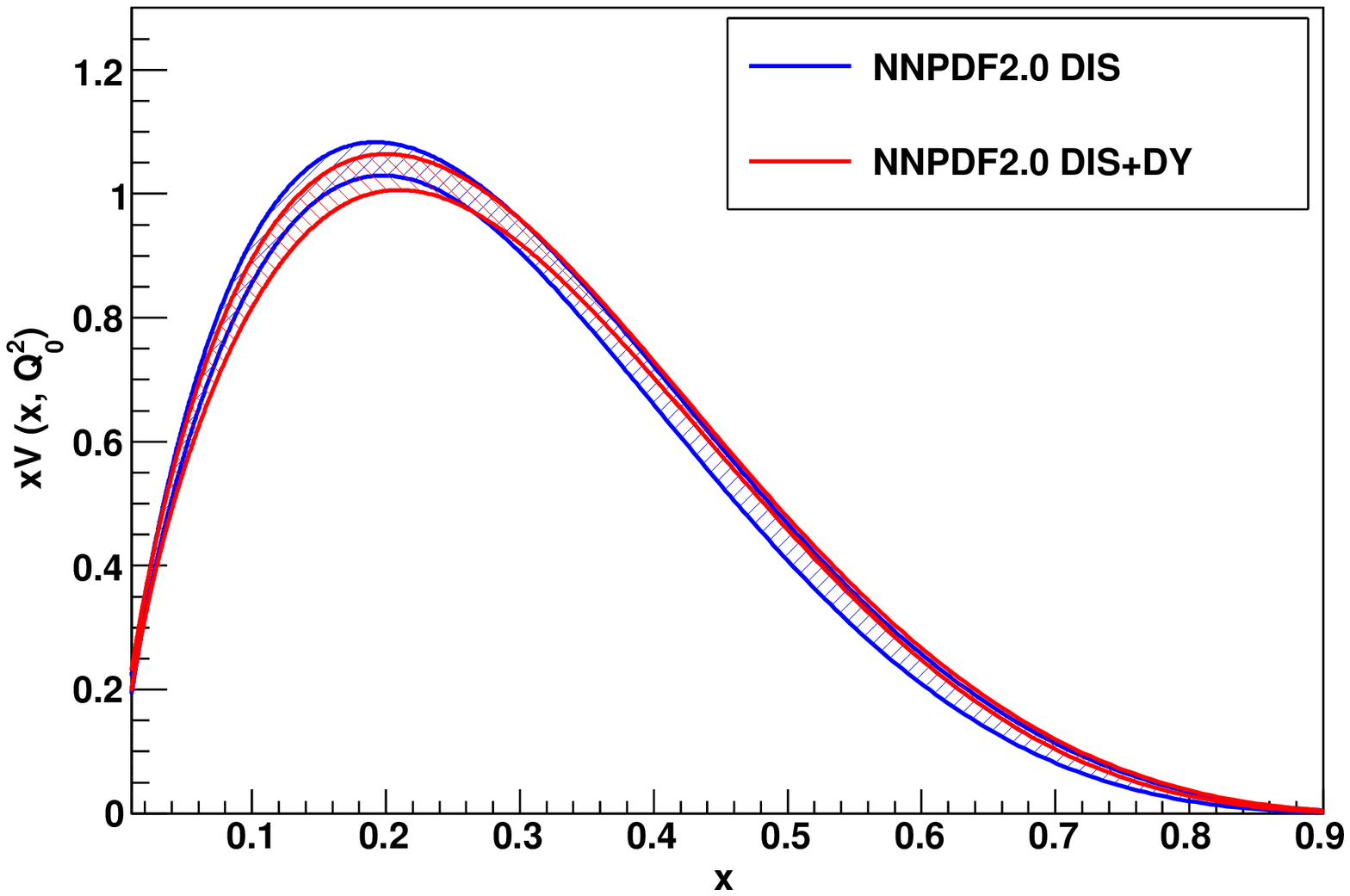}
\includegraphics[width=0.48\textwidth]{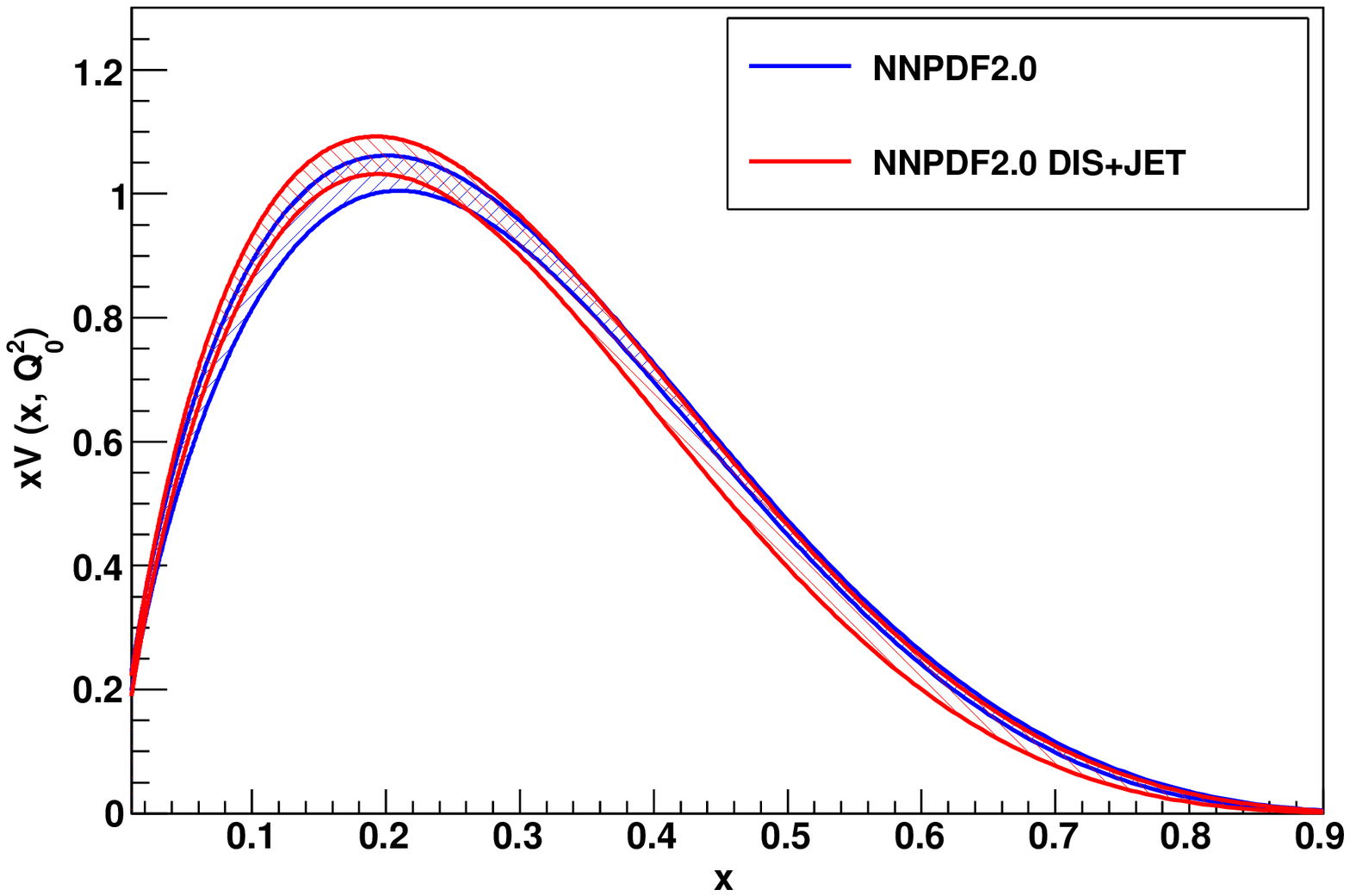}
\includegraphics[width=0.48\textwidth]{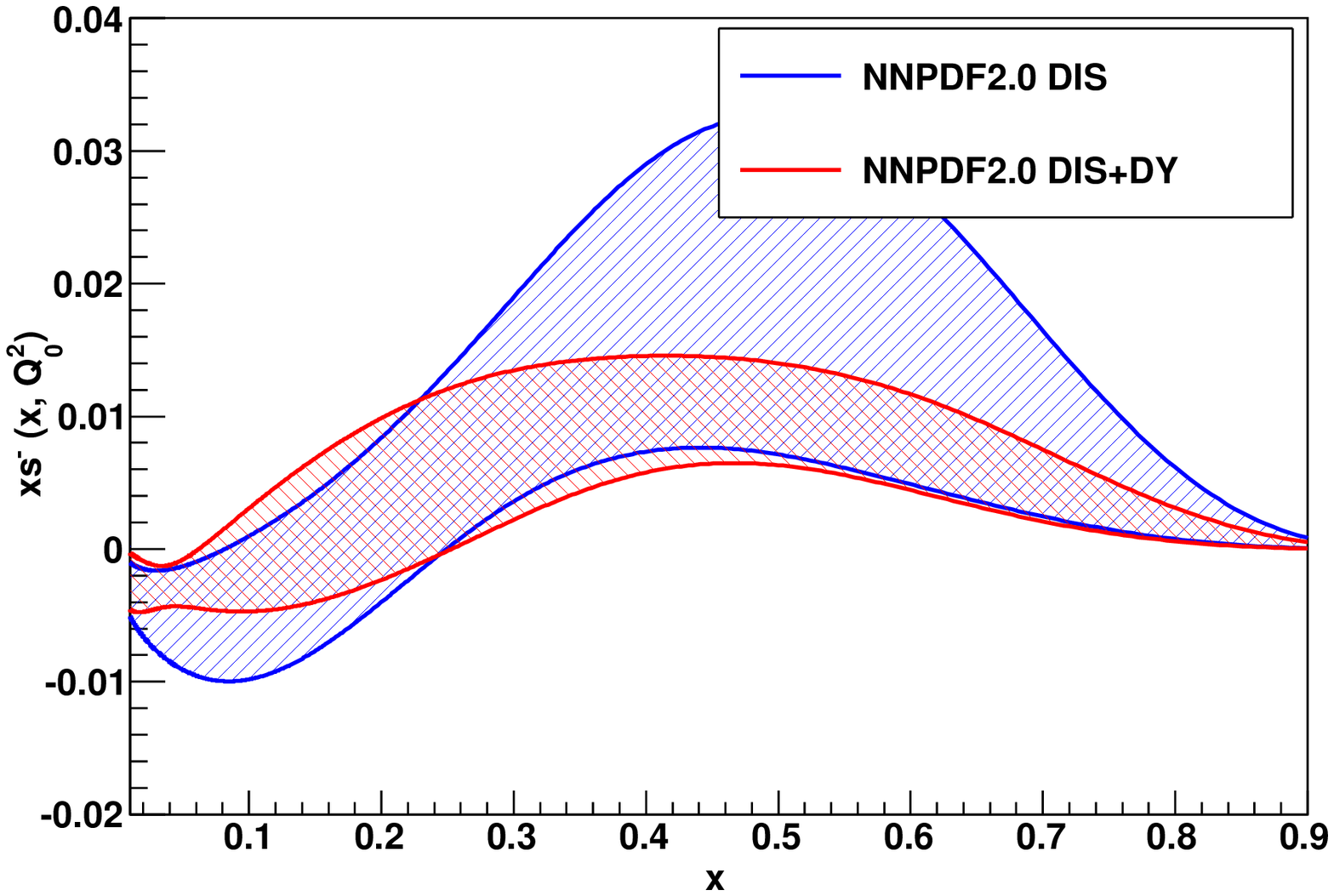}
\includegraphics[width=0.48\textwidth]{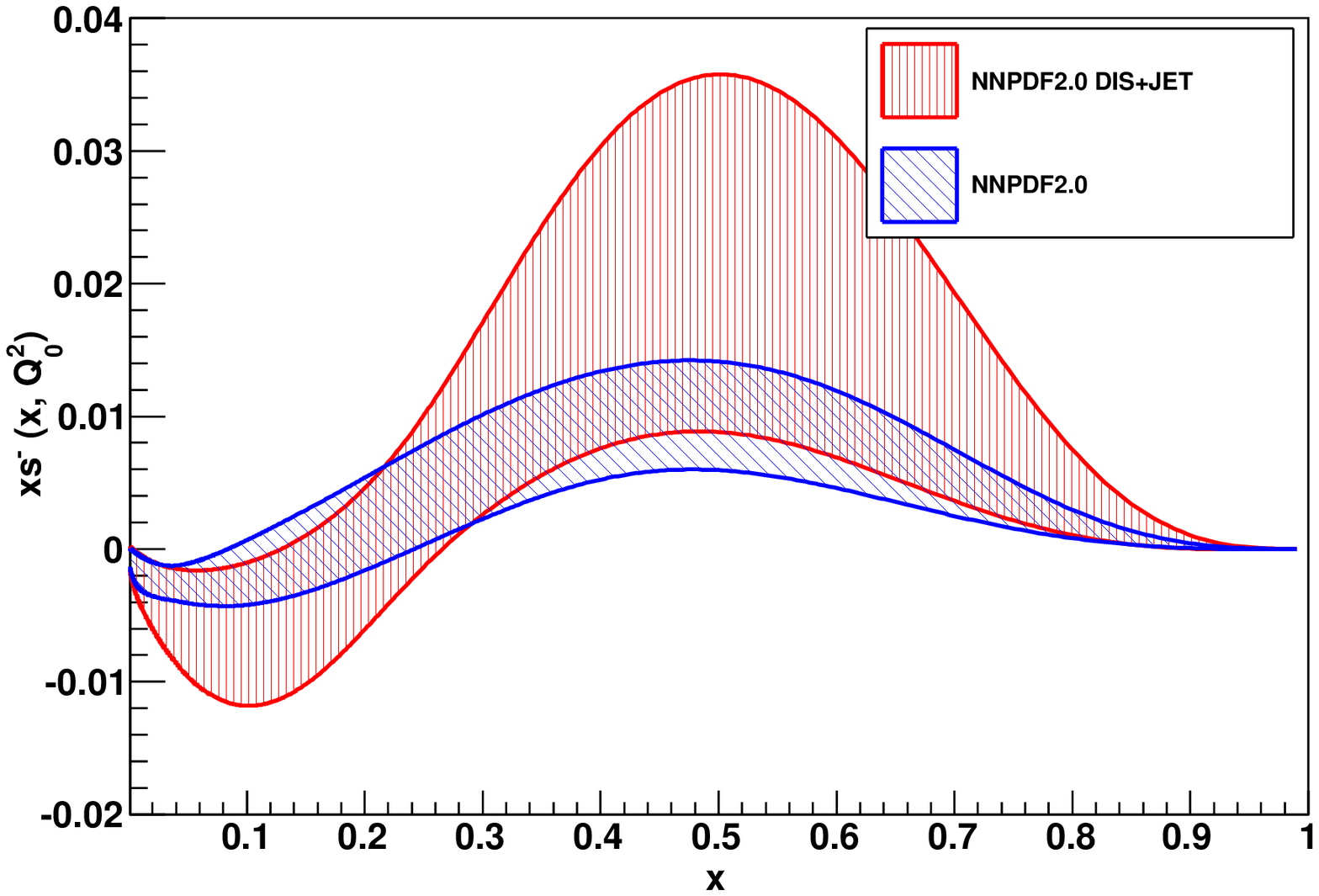}
\caption{ \small Impact of the inclusion of Drell--Yan data in a fit
  with DIS data only (left), and in a fit with DIS and jet data
  (right). From top to bottom, total valence
and  $s-\bar s$ PDFs.
\label{fig:commute}} 
\end{center}
\end{figure}

A more detailed consistency check is performed by comparing fits 
in which a certain "new" dataset is added to different pre-existing
datasets, and verifying that the impact of new data is independent of
the choice of   the dataset to which they are added, thereby also
verifying the mutual consistency of the various data subsets involved.
One such
comparison  (within the framework of
the NNPDF2.0~\cite{Ball:2010de} PDF set) is shown in
Fig.~\ref{fig:commute} in which the effect of 
Drell-Yan data on the total valence and strange valence
PDFs are compared when these data are added to a fit to DIS data only,
or to a fit to DIS+jet data. More tests of this kind were shown in
Ref.~\cite{Forte:2010dt} and demonstrated equally good consistency.

\begin{figure}[t!]
  \begin{center}
    \includegraphics[width=0.45\textwidth]{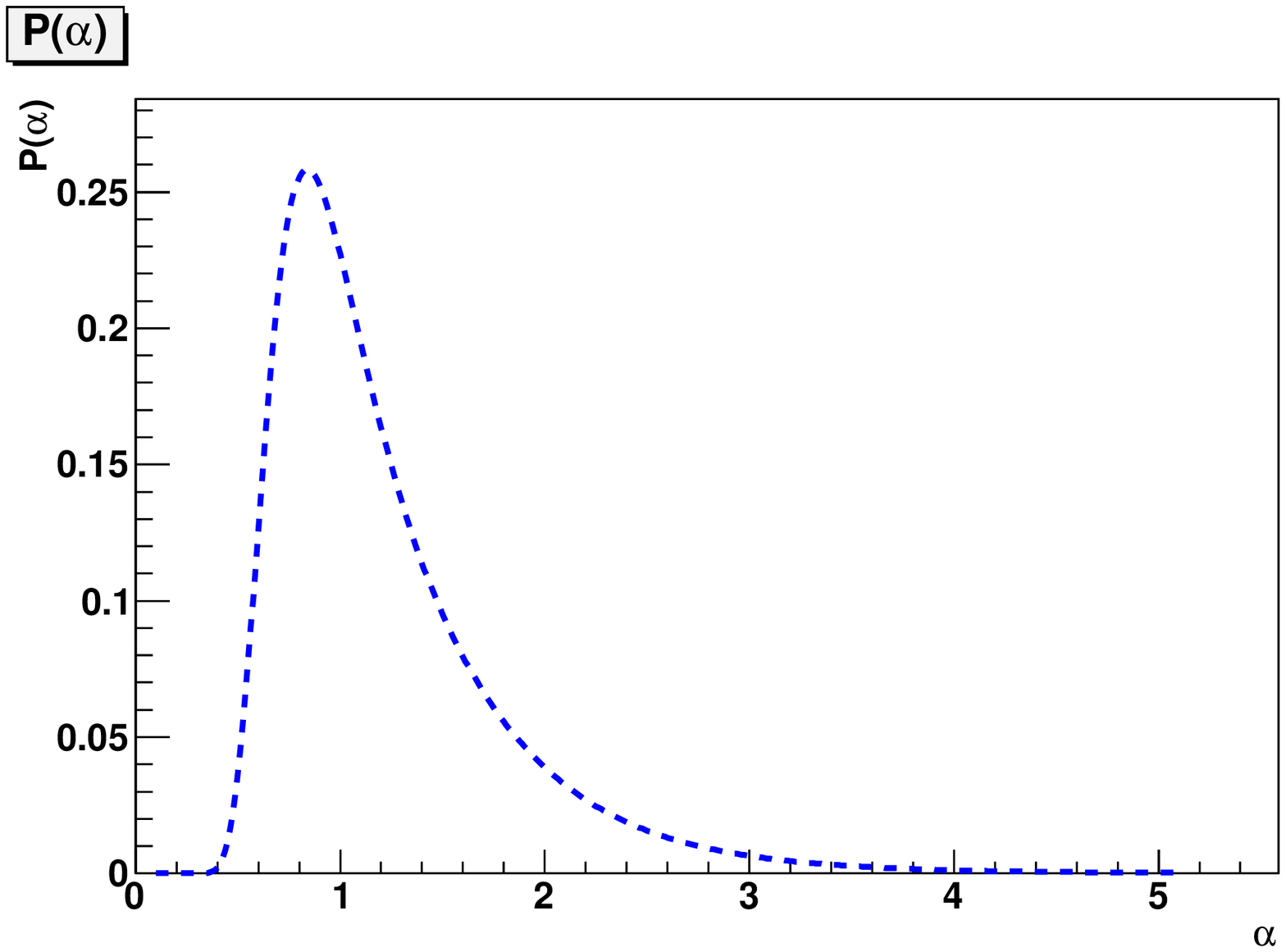}
\includegraphics[width=0.45\textwidth]{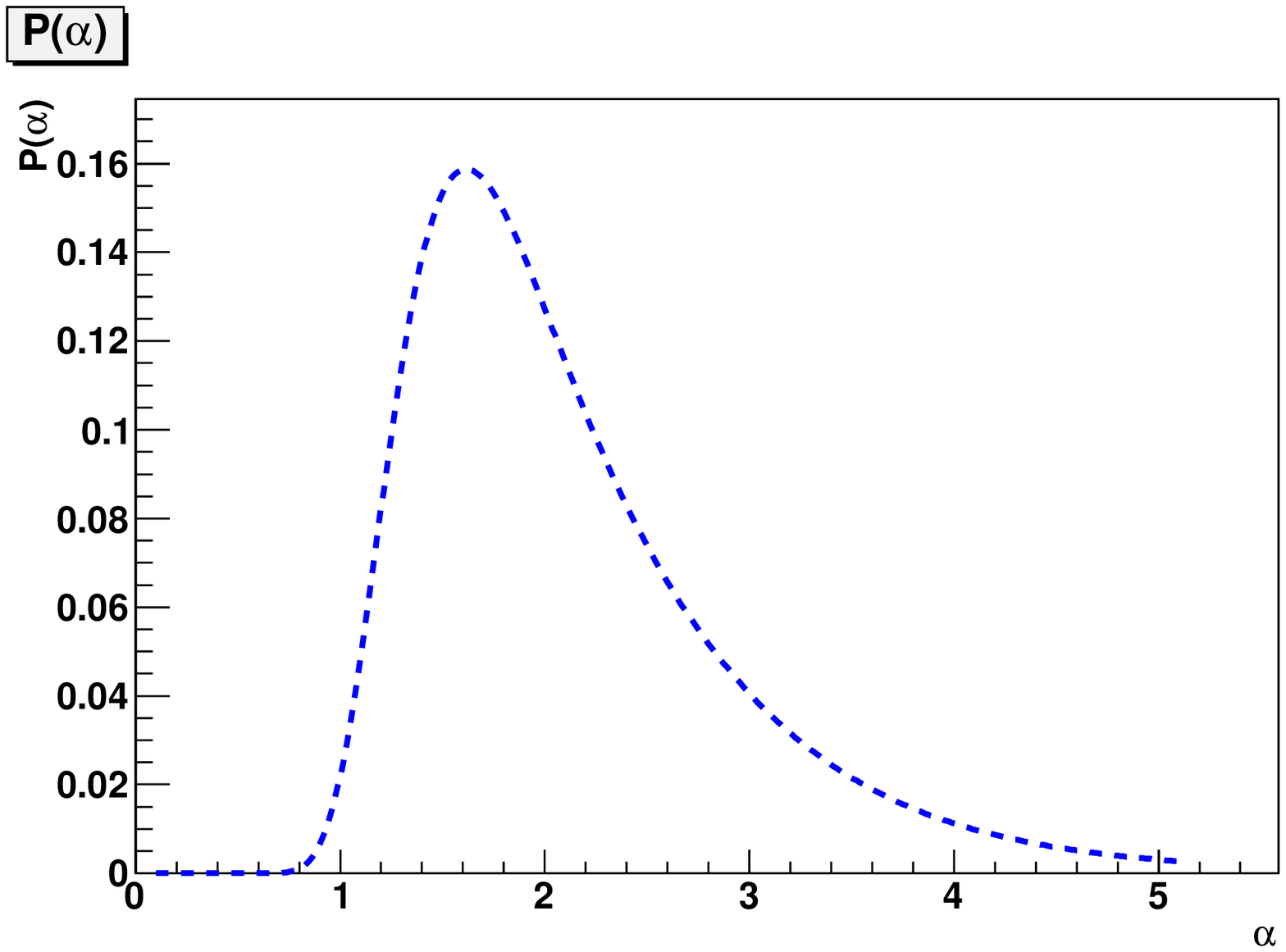}
  \end{center}
  \caption{\small The probability distribution 
    ${\mathcal P}(\alpha)$ for the D0 lepton asymmetry data
uncertainty to be underestimated by a
factor $\alpha$: 
muon data~\cite{d0muon} (left) and electron
data~\cite{Abazov:2008qv} (right). 
\label{fig:chi2-muon} }
\end{figure}

The consistency of different data can  be addressed quantitatively
using the Bayesian reweighting technique of Ref.~\cite{reweighting}
summarized 
in Sect.~\ref{sec:cons}. Namely, assume that the covariance matrix for a given
dataset is rescaled  by a common factor $\alpha$, $\sigma_{ij}\to
\alpha\sigma_{ij}$ so that for that
experiment $\chi^2\to\chi^2/\alpha^2$. It is then easy to
show~\cite{reweighting} that  the probability density
$\mathcal{P}(\alpha)$ for 
$\alpha$ given the data is 
\begin{equation}
\mathcal{P}(\alpha)\propto \smallfrac{1}{\alpha}\sum_{k=1}^N  w_k(\alpha),
\label{eq:rescaling}
\end{equation}
where $w_k(\alpha)$ are the weights \eq{eq:weights} evaluated with the
rescaled covariances.
If $\mathcal{P}(\alpha)$ peaks close to one the new 
data are consistent, while if it peaks far above one, then it is 
likely that the errors in the data have been underestimated.
As an example, we show in Fig.~\ref{fig:chi2-muon}
$\mathcal{P}(\alpha)$ computed for two
of the Tevatron D0 lepton asymmetry
datasets 
analyzed 
in~\cite{reweighting}. For muon data~\cite{d0muon} ${\mathcal P}(\alpha)$
is peaked close to one, implying that
this dataset is consistent with the other
sets in the global fit. For muon data
$\mathcal{P}(\alpha)$ is peaked far from one, suggesting
that experimental uncertainties have been
underestimated by about a factor two.

\section{ Functional and Data
components of the PDF uncertainty}
\label{sec:funcdat}

Because PDFs are functions determined from a finite set of data, one
may expect that on top of the propagated  uncertainty due to the
uncertainty in the data there might be a further uncertainty due to
existence (for sufficiently  general parametrization) of  many PDFs
which give a fit of the same quality to the data. For definiteness, we
will call these different sources of uncertainty
``data'' and ``functional'' uncertainty respectively.
If one were to accept infinitely coarse (e.g. fractal) PDF shapes  the functional
uncertainty would be infinite, but even if it is kept under
control by some smoothness assumption it will generally still be
nonzero. In fact, it was recently argued in Ref.~\cite{Pumplin:2009bb}
that the so called ``tolerance''
criterion~\cite{Pumplin:2002vw}
in
PDF fits which make use of underlying functional
forms with a relatively small number of parameters, and 
 amounts to a rescaling of the
$\Delta \chi^2$ range used
to determine the one--$\sigma$ range,  mostly accounts
for the fact that the choice of a fixed functional form with few
parameters substantially underestimates the functional uncertainty.

\begin{figure}
\centering\includegraphics[width=.48\linewidth]{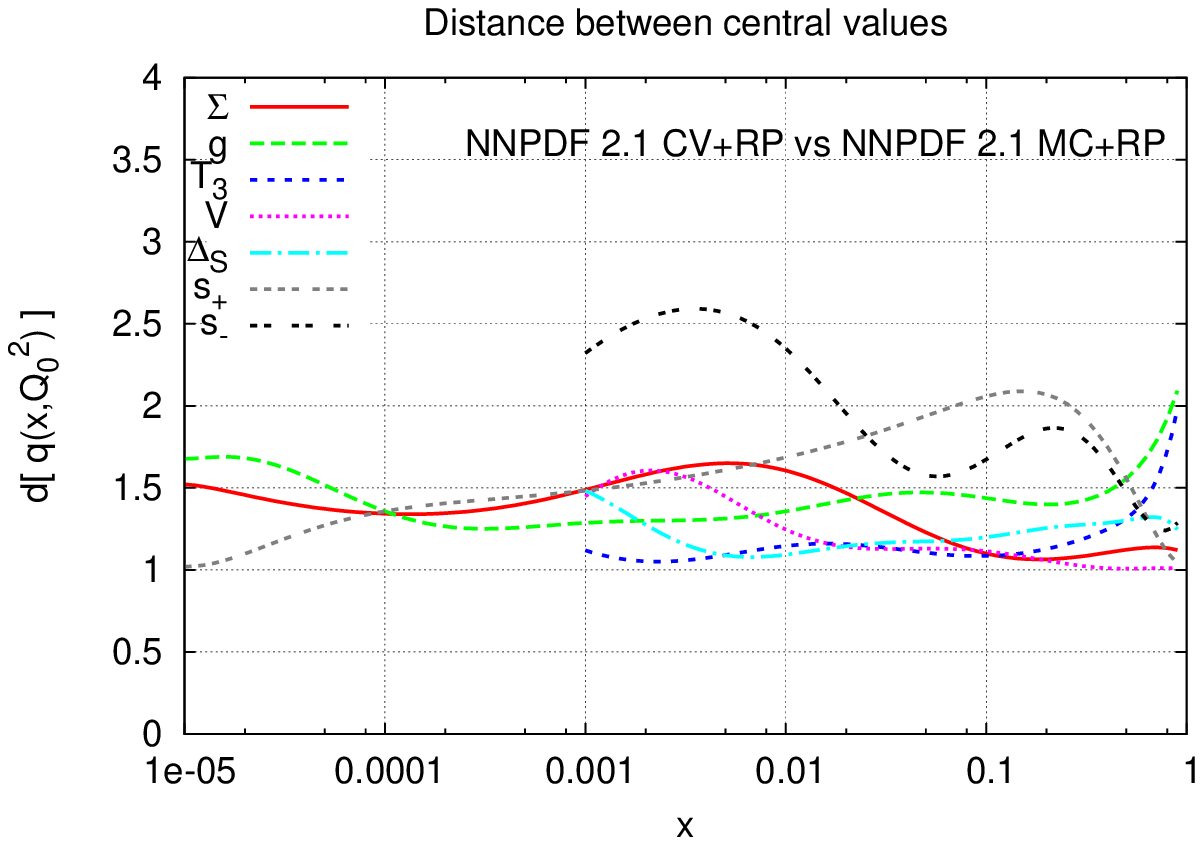}
\centering\includegraphics[width=.48\linewidth]{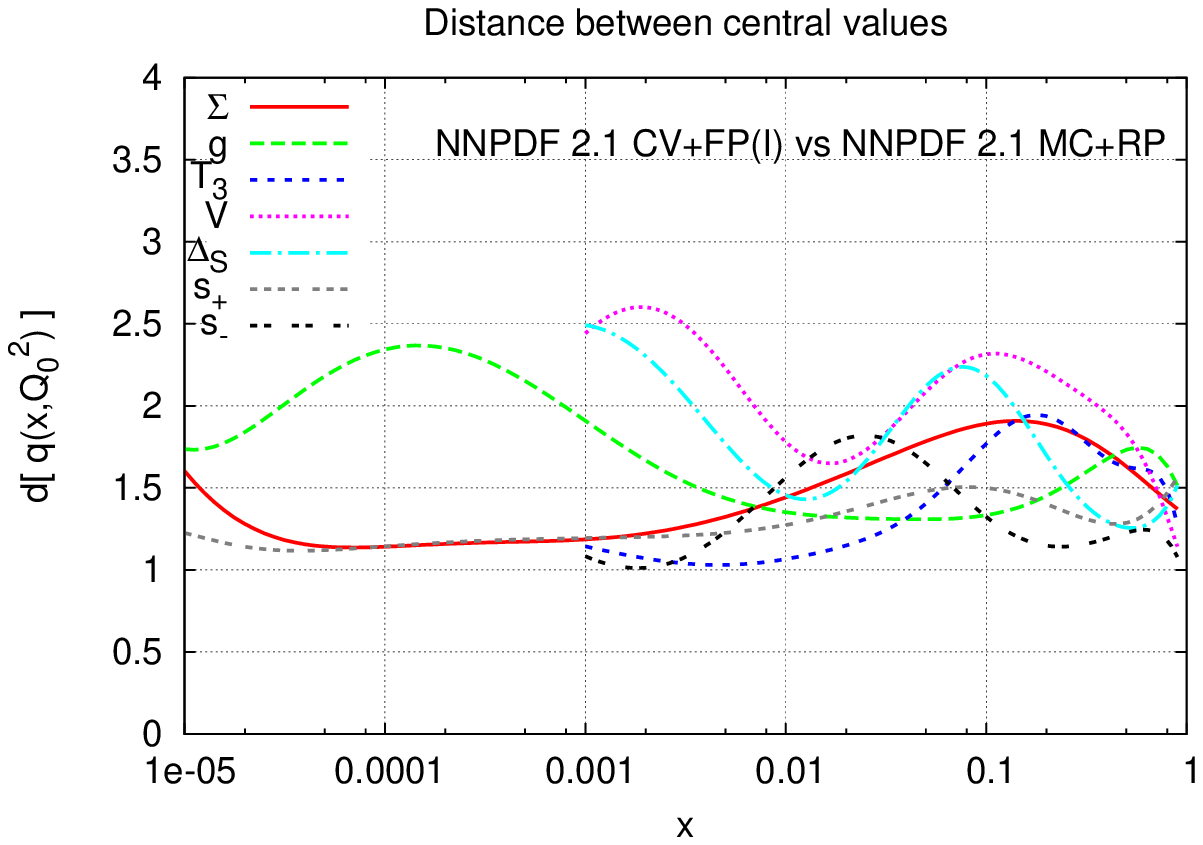}
\caption{Distances between central values
of the reference PDFs and those fitted to different partitions of
central values  (left) or to a fixed partition of central values
(right).
}
\label{fig:distances-nnpdf21-ref-cv-rp}
\end{figure}

In the NNPDF approach, we can actually estimate the relative size of
the data  and functional uncertainty by constructing PDF replica sets
based on a frozen set of data, as we now discuss.
First, we switch off the pseudodata generation. Each PDF replica is
then fitted to the same central data values (CV fit). However, each replica is still
fitted to a different subset of data because for each replica the data
are randomly divided in a training and validation set.
Next, we also switch off the random partitioning of data for each
replica, and we simply fit all PDF replicas to the same partition of
central values (FP). In the latter case, the procedure is repeated
five times, with different choices of the fixed partition in each
case,  in order to make sure that there is nothing special about
the single partition that has been chosen in the first place, and
results are the averaged.

\begin{figure}.
\centering\includegraphics[width=.48\linewidth]{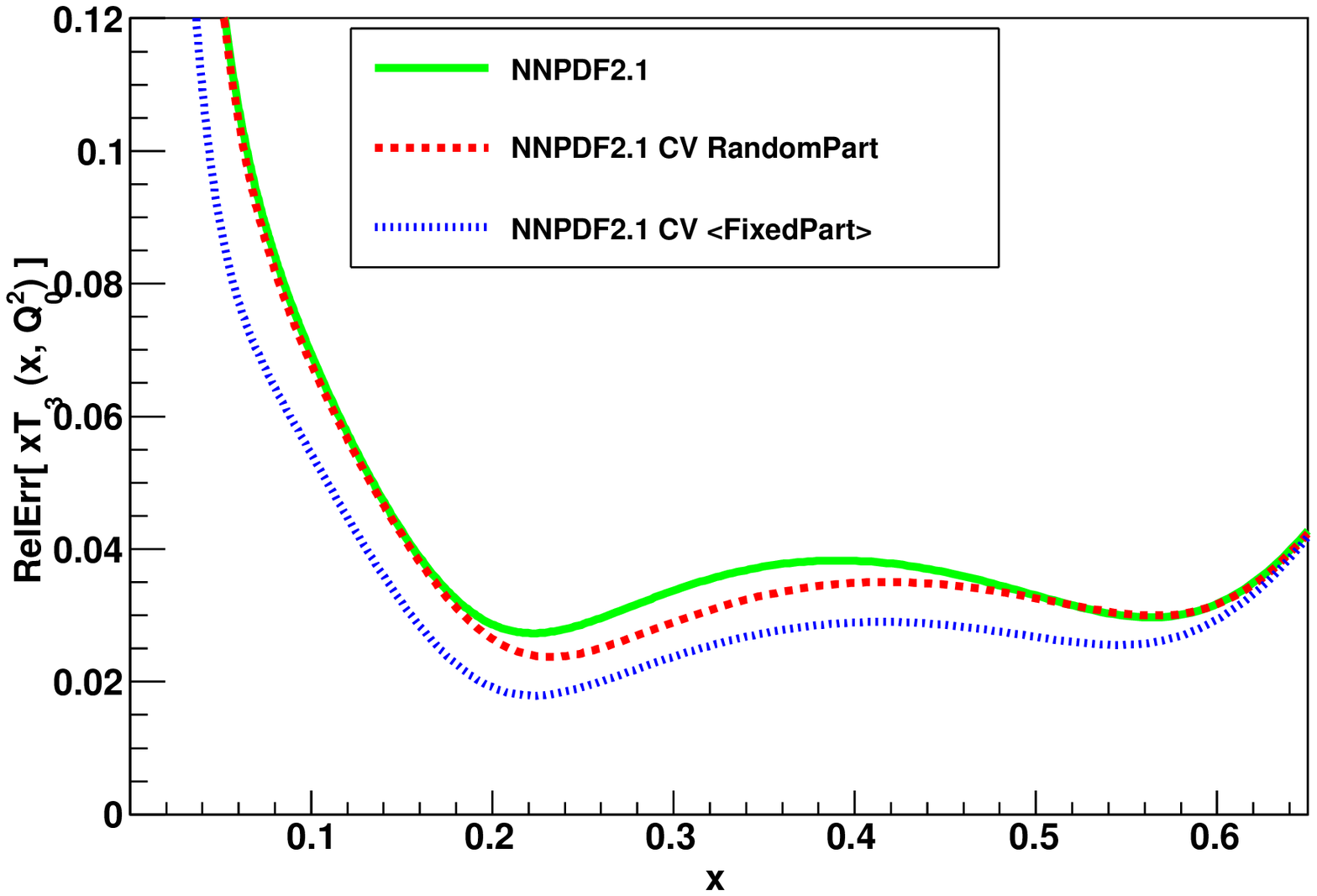}
\centering\includegraphics[width=.48\linewidth]{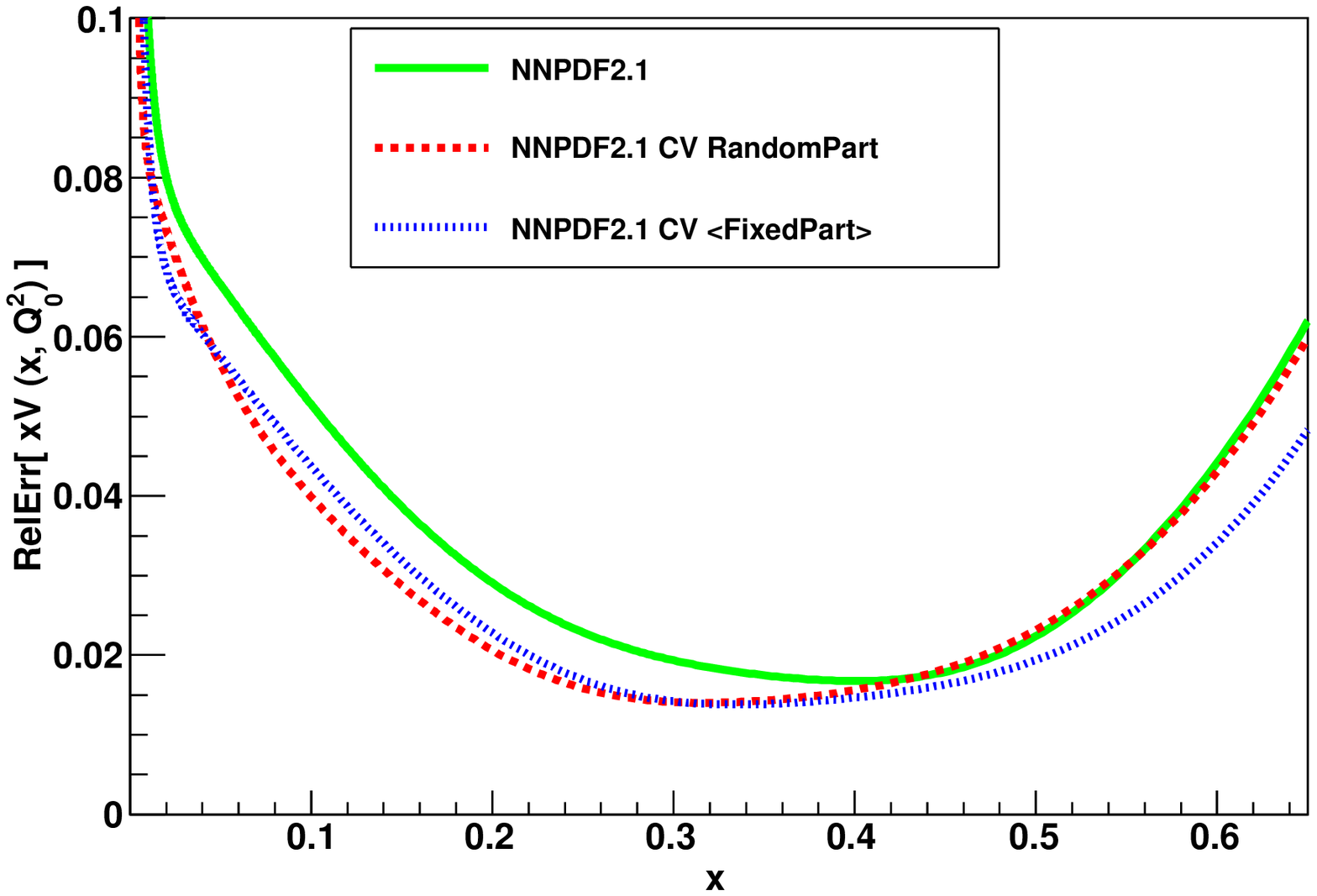}
\caption{Comparison between relative PDF uncertainties of the 
reference PDF set, a fit to varying partitions of central values and a
fit to a fixed partition of central values for the
isotriplet $T_3(x)$
(left) and the total valence $V(x)$ (right).}
\label{fig:pdfplots-partitions}
\end{figure}

Results for the statistical estimators for these fits 
are compared to those of the
default case in Table~\ref{tab:estfit1}. Furthermore, in 
Fig.~\ref{fig:distances-nnpdf21-ref-cv-rp} we display the distances
between central values of PDFs obtained in the various cases, while in
Fig.~\ref{fig:pdfplots-partitions} we compare the relative percentage
uncertainties for a couple representative PDFs.
The central values appear to be very stable (distances of order one)
and indeed the fit quality as measured by $\chi^{2}_{\tot}$ is
essentially the same in all cases. 
When the pseudodata
generation is switched off,  $\la E \ra$, the average quality
of the fit of each replica to the corresponding data replica
now by construction 
coincides with $\la \chi^{2(k)} \ra$ (the same quantity but
computed for central data). Interestingly, the value of 
$\la \chi^{2(k)} \ra$ in the reference and CV fit is identical: this
confirms that the fitting methodology is very efficient in removing
the extra fluctuation of the pseudodata about their central values
induced by the pseudodata generation. It also suggests that the
pseudodata generation is barely necessary. In fact,  one could
take this CV fit as a default: the fluctuations in central data are
then just reproduced by bootstrap, by the process of choosing different
partitions.
Indeed, comparison of PDF uncertainties in the reference and CV case
shows that they are very close and only moderately larger in the
reference case, so that even if the pseudodata generation is viewed as
a more conservative way of estimating uncertainties, in practice it is
seen to have little effect.  

\begin{table}
\begin{center}
\small
\label{tab:nnpdf-fixedparts}
\begin{tabular}{|c|c|c|c|c|c|}
\hline
Dataset  & $\sigma$ Data (\%) & $\sigma$  Ref.
(\%)  &  $\sigma$ CV (\%) &  $\sigma$ FP (\%) \\
\hline
\hline
TOTAL             & 11.3 &      3.7 & 3.8 &  3.1   $\pm$ 0.2 \\
\hline
NMC-pd             & 1.9 &      0.5 & 0.5  & 0.4  $\pm$  0.03 \\
NMC                & 5.0&      1.6 & 1.6 & 1.4  $\pm$  0.2 \\
SLAC               & 4.4&      1.7 &  1.7 & 1.4  $\pm$  0.3 \\
BCDMS              & 5.7 &      2.6 & 2.8 &  2.3  $\pm$  0.3 \\
HERAI-AV           & 2.5&      1.3 & 1.3  & 1.1  $\pm$  0.1 \\
CHORUS             & 15.1&      4.5 & 5.3 & 3.4  $\pm$  0.3 \\
FLH108             & 72.0 &      4.1 & 3.9  & 3.9  $\pm$  0.5 \\
NTVDMN             & 21.1&     14.5 & 14.1 & 12.7  $\pm$  1.6 \\
ZEUS-H2            & 13.4 &      1.3  & 1.3 &  1.1  $\pm$  0.2 \\
ZEUSF2C             & 23.3&     3.1 & 3.1  & 2.8  $\pm$  0.2 \\
H1F2C               & 17.3&     2.9 & 2.9  & 2.6  $\pm$  0.2 \\
DYE605              & 22.3&     8.1 & 7.0  & 6.1  $\pm$  0.3 \\
DYE886              & 20.1&     9.1 & 8.3  & 8.2  $\pm$  0.4 \\
CDFWASY             & 6.0 &     4.5 & 3.4 & 3.1  $\pm$  0.3 \\
CDFZRAP             & 11.5&     3.5 & 3.6 & 3.5  $\pm$  0.5 \\
D0ZRAP              & 10.2&     2.8 & 3.0  & 2.9  $\pm$  0.5 \\
CDFR2KT             & 22.8&     4.8 & 4.4  & 4.4  $\pm$  0.2 \\
D0R2CON              & 16.8 &    5.5 & 5.1 & 5.1  $\pm$  0.2  \\
\hline
\end{tabular}
\caption{The average percentage uncertainty for each
datasets for the reference, central value, and fixed partition PDF
sets.}
\end{center}
\end{table}

However, the most striking result is given by
the PDF uncertainties in the FP case: these
uncertainties, though somewhat smaller, 
are still of the same order of
magnitude as those of the the standard fit. 
This means that different replicas constructed by
refitting exactly the same data over and over again still have a
non-negligible spread and thus uncertainty. This is only possible because of the
random nature of the fitting algorithm, and it shows that indeed there is
a nontrivial space of almost equivalent minima. It should be noticed
that indeed the fluctuation of $\la \chi^{2(k)} \ra$ for this replica
set is significantly smaller than for the reference and CV sets,
consistent with the hypothesis that one is now exploring a space of
equivalent or almost equivalent minima.

A more quantitative insight on the relative size of various
contributions to the uncertainties can be obtained by computing the
average uncertainty on the prediction for the fitted observables
obtained using each PDF set. These are shown, both for the global and
individual dataset, in Table~\ref{tab:nnpdf-fixedparts}, where the
starting data uncertainty  is also shown for comparison.
The uncertainties obtained  fitting to central data or to
pseudodata replicas are almost identical: as already noticed, one might as
well fit to central data. Both are significantly smaller than the
original data uncertainty, thereby showing that an underlying law has
been learnt. The residual uncertainty in the FP case is still sizable. If
one assumes that the uncertainty in the FP case is the functional
uncertainty, while in the CV case it is the sum in quadrature of data
and functional uncertainty, then one concludes that the functional
uncertainty is rather more than half  the total uncertainty.

\section{Outlook}

Having verified that PDFs determined with the NNPDF methodology are
consistent with statistical expectations and free of
parametrization bias, it is natural to think that some of the
statistical tools discussed here, as well as more refined statistical
tests, may be used to guide and validate further
improvements.

Two aspects  of the methodology may be may be amenable to improvement. The
first has to do with the underlying functional form. At present, PDFs
are parametrized as a neural network, multiplied by a preprocessing
function of the form $x^\alpha  (1-x)^\beta$. The exponents are then
randomly varied in a reasonable range. The
preprocessing speeds up the fitting of the neural network, and
ensures that outside the data region the behaviour of the PDF does
not fluctuate too wildly. This procedure is much more general and
unbiased than that used in fits such as MSTW or CTEQ, in which the
functional form also incorporates the same small- and large-$x$
behaviour, but the
exponents $\alpha$ and $\beta$ are fitted (instead of being varied
in a range around their best fit) and the residual number of
parameters is smaller by more than one order of magnitude. But the
preprocessing 
could still be a source of residual bias, so one should check whether results
are stable upon completely different choices of preprocessing. The
second has to do with the determination of the best fit. While 
cross-validation is quite efficient on average, it could still
lead to some specific dataset being under- or overlearnt; it involves
some arbitrariness, for instance in deciding the precise form of the
stopping criteria; and it could lead to an excessively wide and thus
sub-optimal space of minima. Hence alternative methods to determine
the optimal fit should be explored.

Correspondingly, two sets of statistical investigations may be worth pursuing 
in order  to guide and validate these improvements. On the one hand,
it may be interesting to study the form of the probability
distributions of PDF replicas: for instance, this could allow one to
directly address the question of what in a conventional procedure 
 is the $\Delta \chi^2$ range
which correponds to a 68\% confidence interval. On the other hand, it
may be useful to investigate systematically the statistical impact of
each dataset, with the aim of arriving at a full ``closure test'' ---
a proof that there is no information loss in extracting PDFs from
data. These improvements may be useful and even necessary for
precision phenomenology at the LHC.

{\bf\noindent  Acknowledgments:}
We thank G.~Cowan, L.~Lyons and H.~Prosper for discussions
and encouragement. M.U. is supported by the 
Bundesministerium  f\"ur Bildung and Forschung (BmBF) of the Federal 
Republic of Germany (project code 05H09PAE).
This work was partly supported by the Spanish MEC FIS2007-60350 grant.

\bibliographystyle{lesHouches}
\providecommand{\href}[2]{#2}\begingroup\raggedright\endgroup

\end{document}